\documentclass[referee]{aa}
\input epsf
\begin{document}

   \thesaurus{03     
              (11.01.2;  
               11.19.1;  
               11.19.3;  
               11.19.5   
              )}
   \title{The nuclear starburst activity in the Seyfert 2 galaxy NGC 7679}

   \author{Q. S. Gu \inst{1}
           \and
           J. H. Huang\inst{1}
           \and
           J. A. de Diego \inst{2}
           \and
           D. Dultzin-Hacyan \inst{2}
           \and
           S. J. Lei \inst{1}
           \and
           E. Ben\'{\i}tez \inst{2}
           }

   \offprints{Qiusheng Gu, email: qsgu@nju.edu.cn}

   \institute{
            Department of Astronomy, Nanjing University,
            Nanjing 210093, P.R. China
         \and
            Instituto de Astronom\'{\i}a,
            Universidad Nacional Aut\'onoma de M\'exico,
            Apdo Postal 70-264, Mexico D.F. 04510,
            M\'exico
     }

   \date{Received --- ; accepted --- }

   \titlerunning{Nuclear starburst in Seyfert 2 galaxy NGC 7679}

   \authorrunning{Gu, etal.}

   \maketitle

   \begin{abstract}

 We present our recent spectrophotometric results of the infrared
 luminous Seyfert 2 galaxy NGC 7679. We find compelling evidence
 of the presence of a recent nuclear starburst, revealed by the
 observations of a) the spectral features of high order Balmer
 absorption lines, b) the weak equivalent widths for CaIIK
 $\lambda$3933, CN $\lambda$4200, G-band $\lambda$4300 and
 MgIb $\lambda$5173, and c) the suggested UV stellar wind resonance
 lines (N V $\lambda$1240, Si IV $\lambda$1400 and C IV $\lambda$1550)
 in the IUE spectrum. Using the simple stellar population synthesis model,
 we find that in the nuclear 2" $\times$ 2" region, the contributions
 from the old, intermediate and young components are 21.7\%,
 42.9\% and 35.4\%, respectively. These nuclear starburst activities might
 be triggered by the close encounter with NGC 7682, as suggested by recent
 numerical simulations.

      \keywords{
               Galaxies: active ---
               Galaxies: Seyfert ---
               Galaxies: starburst ---
               Galaxies: stellar content
               }
   \end{abstract}

\section{Introduction}

Observational evidence in the sense that the nuclear/circumnuclear
starburst coexists with the central active nucleus has increased
recently. In the study of a sample of Seyfert 2 galaxies,
Storchi-Bergmann et al. (\cite{sb00}) and Gonzalez Delgado et al
(\cite{gd01}) have found that about 30 to 50 \% Seyfert 2 galaxies
have nuclear starbursts. Since both of them might be related to
gas inflow, which could be triggered by the axis-asymmetric
perturbation (such as bars, tidal interactions or mergers), there
is extensive speculation on the connection between starburst and
nuclear activity (Norman \& Scoville \cite{norman88}; Terlevich et
al. \cite{terlevich90}; Heckman et al. \cite{heckman95}; Heckman
et al. \cite{heckman97}; Gonzalez Delgado \cite{gd98}; and see the
recent review by Veilleux \cite{veilleux00}). Gu et al.
(\cite{gu01}) studied 51 Seyfert 2 galaxies from data available in
the literature, and found that Seyfert 2 galaxies with no evidence
of a hidden Seyfert 1 nucleus have characteristics in common with
starburst galaxies.

There are several observable spectral signatures to find the
nuclear starburst in Seyfert 2 galaxies. For example, in the
ultraviolet (UV) band, the stellar wind resonance lines, such as N
V $\lambda$1240, Si IV $\lambda$1400 and C IV $\lambda$1550
(Heckman et al. \cite{heckman95}; Heckman et al. \cite{heckman97};
Gonzalez Delgado et al. \cite{gd98}). In the optical, the broad
emission lines of Wolf-Rayet stars around 4686 \AA (Kunth \&
Contini \cite{kunth99}) and high order Balmer absorption lines
originating in the photosphere of O, B and A stars (Gonzalez
Delgado et al. \cite{gd99}; Gonzalez Delgado et al. \cite{gd00}).
Recently, Cid Fernandes et al. (\cite{cid01}) have suggested a
series of empirical criteria for the existence of starburst in
Seyfert 2 galaxies, such as weak CaIIK $\lambda$3933, low
excitation lines and large far-infrared luminosity, etc.

In this paper, we report the presence of spectral signatures of
nuclear starburst in the infrared luminous Seyfert 2 galaxy NGC
7679, with the infrared luminosity, $L_{IR}$, of $1.1 \times
10^{11} \rm L_\odot$(Veilleux et al. \cite{veilleux95} ;
Veron-Cetty \& Veron \cite{vcv}) and the redshift ({\it{z}})
0.01714 (corresponding to a distance of 68.5 Mpc, for H$_0$ = 75
km s$^{-1}$ Mpc$^{-1}$ and q$_0$ = 0). NGC 7679 is one member of
Arp 216, the other is NGC 7682, their projected separation is
89.6kpc (Arp \cite{arp}). We suggest that the starburst activity
in this galaxy might be related to the tidal interaction with NGC
7682, as indicated by recent numerical simulations (Barnes \&
Hernquist \cite{barnes96}; Mihos \& Hernquist \cite{mihos96}).
This paper is organized as follows. In Sect.2 we describe our
observations, data reduction and results. In Sect. 3 we discuss
our results and summarize our conclusions in Sec. 4.

\section{Observations and results}

Spectra of NGC 7679 were obtained on Sept. 28-29, 2000, with the
Boller \& Chivens spectrograph attached to the 2.1m telescope of
Observatorio Astron\'{o}mico Nacional at San Pedro Martir
(Mexico). We used a Thomson 2048 $\times$ 2048 CCD and a 600
lines/mm grating, which provides a dispersion of 83 \AA/mm. The
width of the long slit was set to 2" (i.e., 664 pc for NGC 7679).
HeAr-spectra had been taken before and after the object spectra
for wavelength calibration, and BD +28 4211 was selected from KPNO
standards for flux calibration. The total exposure time was 10,800
s (2,700s $\times$ 4). All spectra were reduced using standard
IRAF \footnote{IRAF is distributed by the National Optical
Astronomy Observatory, which is operated by the Association of
Universities for Research in Astronomy, Inc., under cooperative
agreement with the National Science Foundation.} procedures.

In Fig. 1, we show the central 2" $\times$ 2" region spectrum of
NGC 7679, and in Table 1, we list all emission and absorption
lines that we identify in the spectrum along with their respective
equivalent widths (EWs). Stellar absorption wings clearly evolve
the H$\beta$ and H$\gamma$ emission lines. For these lines, fluxes
and EWs were obtained using the IRAF-\textit{specfit} routine. The
spectrum also shows the presence of high order Balmer absorption
lines(up to H12), and several weak absorption lines, such as,
CaIIK $\lambda$3933, CN $\lambda$4200, G-band $\lambda$4300 and
MgIb $\lambda$5173.

\begin{table*}
\caption{Absorption and emission lines in NGC 7679}
\begin{tabular}{llcc}
\hline \hline
Wavelength (\AA) & Ion & Equivalent Width (\AA)\\
\hline
3727 & [O II]    & 16.33$^{e}$ \\
3750 & H12       & 2.47 \\
3770 & H11       & 4.36 \\
3797 & H10       & 5.32 \\
3835 & H9        & 6.62 \\
3889 & H$\zeta$  & 7.25 \\
3933 & CaII K    & 1.41 \\
3970 & H$\epsilon$& 6.73 \\
4101 & H$\delta$ & 6.79 \\
4200 & CN band   & 0.35 \\
4300 & G band    & 0.68 \\
4340 & H$\gamma$ & 2.92$^{a,e}$\\
4861 & H$\beta$  & 9.72$^{b,e}$\\
4959 & [O III]   & 3.42$^{e}$ \\
5007 & [O III]   & 9.74$^{e}$ \\
5173 & MgIb      & 0.82 \\ \hline
\end{tabular}

\noindent $^{a}$ Corrected for stellar absorption of 6.0 \AA \
using \textit{specfit}.

\noindent $^{b}$ Corrected for absorption of 5.7 \AA.

\noindent $^{e}$ Means for emission lines.
\end{table*}

\begin{figure}
\vspace{0cm} \epsfxsize=17cm \centerline{\epsfbox{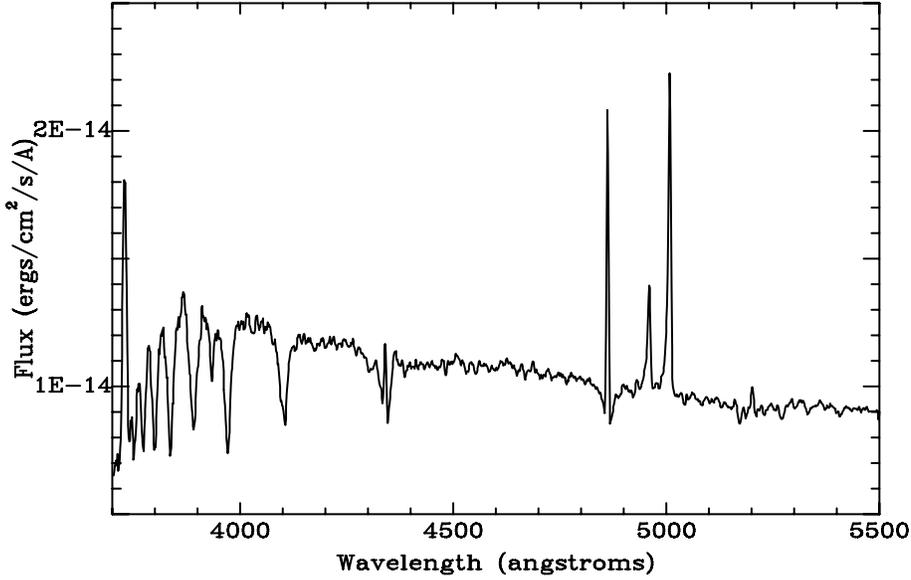}}
\vspace{0cm} \caption{The spectrum of NGC 7679, where we could see
the clear higher order Balmer absorption lines.}
\end{figure}

\section{Discussion}

\subsection{Stellar populations}
To estimate the stellar population in the nuclear region of NGC
7679, we compare our observed spectrum with the evolutionary
synthesis model, GISSEL96 (Bruzual \& Charlot \cite{bc96}).
Following Cid Fernandes et al.(\cite{cid01}), we use three
components: an old and an intermediate populations, with ages $>$
1 Gyr and $\sim 10^8$ yr, respectively, and an $f_{\lambda} \sim
\lambda ^{-1}$ power-law component to account for either a
scattered AGN continuum or a young starburst component, since it
is impossible to disentangle the power-law AGN continuum from the
contribution of a young starburst of an age $\leq$ 10 Myr
(Storchi-Bergmann et al. \cite{sb00}; Gonzalez Delgado et al.
\cite{gd01} and Cid Fernandes et al. \cite{cid01}).

We normalized all spectra at 4800 \AA, and we compare parameters
of continuum fluxes at 4020, 4510 and 5500 \AA, as well as EWs of
the absorption lines for CaIIK $\lambda$3933,CN $\lambda$4200 and
G-band $\lambda$4300. The relative errors on these quantities are
set to less than 5\%. Fig. 2 shows the observed spectrum and the
three component population synthesis fit. The fractions of the
power-law component, the intermediate and the old populations
contributing to the total 4800 \AA \ monochromatic light are
35.4\%, 42.9\% and 21.7\%, respectively.

For comparison, we make use of the empirical formulae by Cid
Fernandes et al.(\cite{cid01}) to derive the fractions of old and
young population, the results are for the young, intermediate and
old population, the fractions are 33.5\%, 48.3\% and 18.2\%,
respectively, which are consistent with our fitting results.

\begin{figure}
\vspace{0cm} \epsfxsize=17cm \centerline{\epsfbox{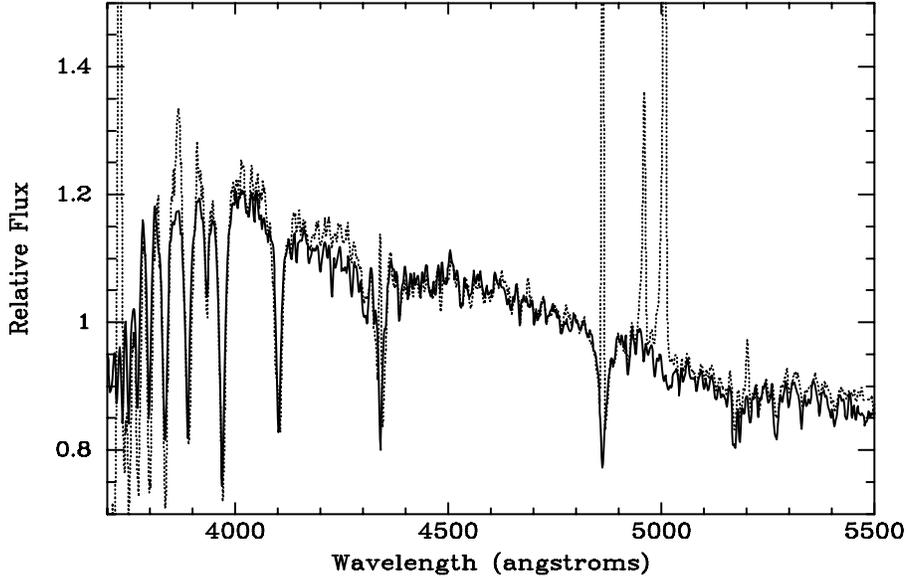}}
\vspace{0cm} \caption{The best-fitting spectrum for NGC 7679 with
three-component population synthesis model(solid line), which is
obtained by combining a power-law component with an intermediate
age($10^8$ yr) and an old ($>$1Gyr) population. The relative
contributions to the total 4800 \AA \ monochromatic light are
35.4\%, 42.9\% and 21.7\%, respectively. For comparison, the
observed spectrum is superimposed with the dotted line.}
\end{figure}

\subsection{UV spectroscopy}

The presence of high order Balmer absorption lines indicate
clearly the existence of intermediate-age stars in the nuclear
region of NGC 7679. But the origin of the power-law component,
accounting for about 35\% of the total 4800 \AA \ flux, may be the
result of a young starburst or a scattered AGN continuum.

It is well known that the stellar wind resonance absorption lines
(N V $\lambda$1240, Si IV $\lambda$1400 and C IV $\lambda$1550)
are unambiguous signatures of young starbursts in the UV band, as
suggested by Heckman et al. (\cite{heckman95}); Heckman et al.
(\cite{heckman97}); and Gonzalez Delgado et al. (\cite{gd98}).
Unfortunately, these features can only be observed clearly in the
brightest UV galaxies(Heckman et al. \cite{heckman97}; Gonzalez
Delgado et al. \cite{gd98}).

Is there any indication of young starburst in the UV spectrum of
NGC 7679 ? We retrieved the UV spectrum for this object from the
IUE public archive (Fig. 3). For comparison, we also plot the IUE
and HST GHRS spectra of NGC 5135. NGC 5135 is one of Seyfert 2
galaxies that presents clear nuclear starburst activity and high
order Balmer absorption lines (Gonzalez Delgado et al.
\cite{gd98}).

Comparing IUE with HST GHRS spectra of NGC 5135, we checked that
all typical absorption lines(as marked in Fig. 3) in high
resolution and high S/N GHRS spectrum are also present in the IUE
data, though the S/N is roughly low. At the same time, it is clear
that all these stellar wind absorption lines are also present in
NGC 7679, in particular, we could see the clear P-Cygni profile of
C IV $\lambda$1550, which is even more significant than that in
NGC 5135. Although the IUE spectrum is marginal, we find the clear
signature of a young starburst in the nuclear region of NGC 7679.

That the power-law component is the contribution of young
starburst found in this work for NGC~7679 is consistent with a
recent work by Cid Fernandes et al.(\cite{cid01}). These authors
have found that all $x_{FC} \geq 30 \%$ sources are confirmed
Seyfert 2/starburst composites (where $x_{FC}$ means the
contribution from a power-law featureless continuum) and in these
sources, the featureless continuum is dominated by
nuclear/circumnuclear starburst, because in Seyfert 2 galaxies the
scattered light from the central AGN cannot exceed $\sim$ 30 \%,
otherwise we should observe the reflected broad lines directly and
the galaxy would be no longer classified as a Seyfert 2 galaxy.

\begin{figure}
\vspace{0cm} \epsfxsize=17cm \centerline{\epsfbox{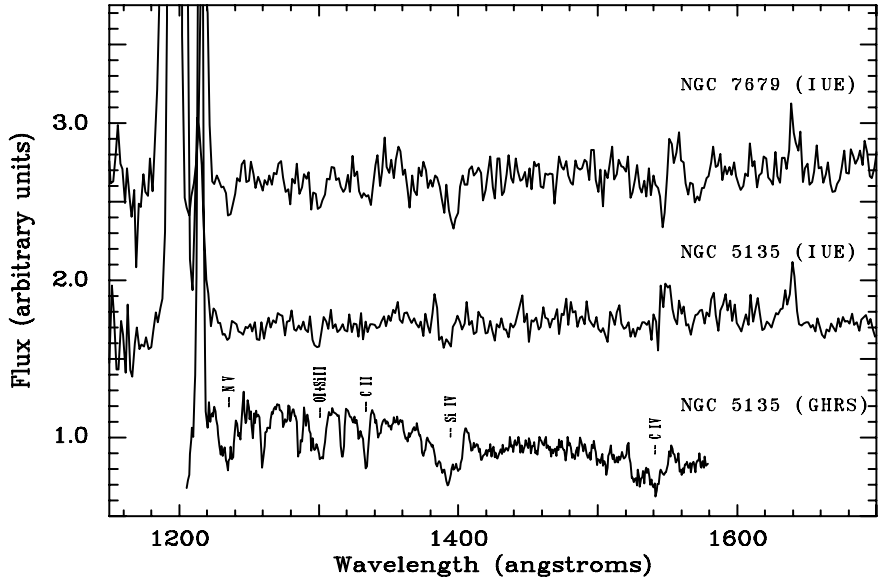}}
\vspace{0cm} \caption{Comparison the IUE UV spectrum of NGC 7679
with NGC 5135, and IUE with HST GHRS for NGC 5135. Though the IUE
spectrum of NGC 5135 is somewhat low S/N, all these marginal
absorption lines are clearly present in the GHRS spectrum. Several
stellar wind absorption lines, such as N V $\lambda$ 1240, Si IV
$\lambda$ 1400 and C IV $\lambda$ 1550, can also be identified in
the IUE spectrum of NGC 7679. }
\end{figure}

\section{Conclusions}

In this paper, we present the unambiguous evidence for the recent
starburst activity in the nuclear region of NGC 7679, which are:
the higher order Balmer absorption lines and weaker CaIIK
$\lambda$3933, CN $\lambda$4200, G-band $\lambda$4300 and MgIb
$\lambda$5173; and in the UV band, several stellar wind absorption
lines. Using the simple three-component stellar population
synthesis model, we obtain the nuclear stellar population could be
42.9\% intermediate age($10^8$ yr), 21.7\% old ($>$1Gyr) combining
with a 35.4\% power-law component, which might be from the young
massive nuclear starburst activity.

\begin{acknowledgements}
 We would like to thank Dr. Rosa Gonzalez Delgado for sending us their
 HST GHRS spectra.
 QSGU acknowledges support from UNAM postdoctoral program (Mexico) and
 support from National Natural Science Foundation of
 China and the National Major
 Project for Basic Research of the State Scientific Commission of China.
 And DD-H, JAD and EB acknowledge support from grant IN 115599 from
 PAPIIT-UNAM.
 This research has made use of NASA's Astrophysics Data System Abstract
 Service and the NASA/IPAC Extragalactic Database (NED) which is operated by
 the Jet Propulsion Laboratory, California Institute of Technology, under
 contract with the National Aeronautics and Space Administration.
\end{acknowledgements}

\end{document}